\documentstyle[12pt]{article}

\newcommand{\deff}[2]{\newcommand{#1}{#2}}
\deff\lform{\hbox{$\sqcup$}\llap{\hbox{$\sqcap$}}}
\deff\half{{\textstyle{1\over2}}} \deff\al{\alpha} \deff\be{\beta}
\deff\ga{\gamma}  \deff\Ga{\Gamma} \deff\de{\delta}  \deff\De{\Delta}
\deff\ep{\epsilon}  \deff\vep{\varepsilon} \deff\ze{\zeta}
\deff\et{\eta} \deff\th{\theta}  \deff\Th{\Theta}  \deff\vth{\vartheta}
\deff\io{\iota} \deff\ka{\kappa} \deff\la{\lambda} \deff\La{\Lambda}
\deff\rh{\rho} \deff\si{\sigma}  \deff\Si{\Sigma} \deff\ta{\tau}
\deff\up{\upsilon}  \deff\Up{\Upsilon} \deff\ph{\phi}  \deff\Ph{\Phi}
\deff\vph{\varphi} \deff\ch{\chi} \deff\ps{\psi}  \deff\Ps{\Psi}
\deff\om{\omega}  \deff\Om{\Omega} \deff\bC{{\bf C}} \deff\bN{{\bf N}}
\deff\bQ{{\bf Q}} \deff\bR{{\bf R}} \deff\bZ{{\bf Z}}
\deff\bGa{{\bf\Ga}} \deff\bLa{{\bf\La}} \deff\cA{{\cal A}}
\deff\cB{{\cal B}} \deff\cF{{\cal F}} \deff\cI{{\cal I}} \deff\cL{{\cal
L}} \deff\cM{{\cal M}} \deff\cN{{\cal N}} \deff\cO{{\cal O}}
\deff\cU{{\cal U}} \deff\cV{{\cal V}} \deff\cW{{\cal W}} \deff\cZ{{\cal
Z}} \deff\KMa{Kac-Moody algebra} \deff\hwm{highest weight module}
\deff\eca{extended conformal algebra} \deff\cft{conformal field theory}
\deff\cfts{conformal field theories} \deff\dCo{{\rm h}^{\vee}}
\deff\hb{\hfill\break} \deff\ie{{\it i.e.\ }} \deff\eg{{\it e.g.\ }}
\deff\proof{\noindent {\it Proof:}\ } \deff\example{\noindent {\it
Example:}\ } \deff\ZZ{Z\!\!\!Z} \deff\NN{I\!\!N} \deff\RR{I\!\!R}
\deff\CC{I\!\!\!\!C} \deff\QQ{I\!\!\!\!Q} \deff\bfg{{\bf g}}
\deff\bfh{{\bf h}} \deff\bfnm{{\bf n}_-} \deff\bfnp{{\bf n}_+}
\deff\hg{\hat{\bf g}} \deff\bft{{\bf t}}
\deff\hz{\hat{z}} \deff\bs{\bigskip}
\deff\hr{\bs\centerline{\vbox{\hrule width5cm}}\bs} \deff\Homg{{\rm
Hom\,}_{\cU(\bfg)}} \deff\Homnm{{\rm Hom\,}_{\cU(\bfnm)}}
\deff\Homnp{{\rm Hom\,}_{\cU(\bfnp)}} \deff\vacL{|\La\rangle}
\deff\vacLp{|\La'\rangle} \deff\aliv{\al_i^\vee} \deff\alp{\al_+}
\deff\hch{\hat{\ch}} \deff\tom{\tilde{\om}} \deff\tch{\tilde{\ch}}
\deff\tph{\tilde{\ph}} \deff\hph{\hat{\ph}}

\newcommand{\bref}[2]{\bibitem{#1}{#2}}
\deff\bn{{\bf n}} \deff\bg{{\bf g}} \deff\bt{{\bf t}}
\deff\bh{{\bf h}}
\deff\ba{{\bf a}} \deff\bb{{\bf b}}
\newcommand{\mb}[1]{\mbox{\boldmath $#1$}}
\deff\HS{Hochschild-Serre}
\deff\bmp{{\bf m_+}} \deff\bmm{{\bf m_-}}
\newcommand{\scm}[1]{H^{\infty/2+#1}}
\deff\cC{{\cal C}}
\def\pigh{\pi_{gh}{}}
\def\bull{\bullet}
\def\Vir{{\bf Vir}} \def\hgt{{\rm ht\,}} \def\deg{{\rm deg\,}}

\def\wbg{{\widehat\bg}} \def\wbh{{\widehat\bh}} \def\wbn{{\widehat\bn}}
\def\wbt{{\widehat\bt}}
\def\sign{{\rm sign}} \def\Ep{{\cal E}} 
\def\pbv{{\bf Vir}}

\newcommand{\eqn}[2]{\begin{equation} #2 \label{#1} \end{equation}}

\newcommand{\thm}[3]{\begin{#1} #3 \label{#1} \end{#1}}
\begin{document}
\bibliographystyle{unsrt}

\renewcommand{\thefootnote}{\fnsymbol{footnote}}
\begin{titlepage}
\begin{flushright}
CERN-TH.6646/92
\end{flushright}
\vspace{1cm}

\begin{center} {\huge\bf
Semi-infinite cohomology in conformal\\ [.2cm]
field theory
and 2d gravity\footnote{To be published in the proceedings
of the XXV Karpacz Winter School of Theoretical Physics,
Karpacz, 17--27 February 1992.}}\\ [1cm]

Peter Bouwknegt \\
{\it CERN - Theory Division} \\
{\it CH-1211 Geneva 23} \\
{\it Switzerland} \\ [.3cm]

Jim McCarthy \\
{\it Department of Physics and Mathematical Physics } \\
{\it University of Adelaide} \\
{\it Adelaide, SA 5001, Australia} \\ [.3cm]

Krzysztof Pilch\footnote{Supported
in part by the Department of Energy Contract \#DE-FG03-91ER-40168.}\\
{\it Department of Physics and Astronomy} \\
{\it University of Southern California} \\
{\it Los Angeles, CA 90089, USA} \\ [1.5cm]

{\bf Abstract}\\
\end{center}

We discuss various techniques  for computing the
semi-infinite cohomology of
highest weight modules which arise in the  BRST
quantization of two dimensional field theories. In particular, we
concentrate on two such theories -- the $G/H$ coset models
and $2d$ gravity coupled to $c\leq 1$ conformal matter.

\vfill
\begin{flushleft}
CERN-TH.6646/92\\ADP-92-194/M12\\USC-92/020\\hepth-th/9209034\\September 1992
\end{flushleft}

\end{titlepage}

\renewcommand{\thefootnote}{\arabic{footnote}}
\addtocounter{footnote}{-2}

\section{Introduction}

  Among the interesting new phenomena discovered in recent studies of
two dimensional gravity models is the existence of so-called discrete
states \cite{Pol,Gross} in the physical spectrum.  Models with a $c\leq
1$ conformal matter sector have been extensively studied in the
framework of BRST quantization, where the physical states represent
nontrivial cohomology classes of the BRST charge -- the discrete states
are then classes present at only a countable set of momenta (see
\cite{LZclo,LZceo,BMPcmpt,LZcmpt,Witsr,WiZw} and references therein).
Perhaps even more striking is the fact that they occur over a range of
different ghosts numbers, quite unlike the physical spectrum of
critical string theories studied previously.

Heuristically, this novel behaviour for the BRST cohomology arises from
two distinct sources: first is the obvious difference in that the
world-sheet scalar fields have background charges for the non-critical
models (the corresponding Fock space is called a Feigin-Fuchs module);
second, in the case of $c<1$ matter, one must project from the Fock
spaces onto the irreducible modules of the Virasoro algebra which
comprise the spectrum of such models. This projection can be
efficiently carried out \cite{BMPcmpt,LZcmpt} using free field
resolutions of the irreducible module in terms of Feigin-Fuchs modules
\cite{Felder}. Thus, both sources are somewhat related, and in all
cases the problem  of identifying the physical states of the theory can
be reduced to the computation of the BRST cohomology of a tensor
product of two Feigin-Fuchs modules.

Mathematically, the BRST cohomology discussed above is an example of
semi-infinite cohomology \cite{Feigin} of the Virasoro algebra.
Another class of algebras whose semi-infinite cohomology has been well
discussed is the affine Kac-Moody algebras \cite{Feigin,FGZ}.  It is
natural, then, to look for the physical setting of the corresponding
BRST complex by analogy with that in $2d$ gravity.   An immediate
example is in the context of the representation theory of affine
Kac-Moody algebras, a problem which is directly relevant to treatments
of general $G/H$ coset models \cite{BaHa,GKO}.  In fact, when
formulated as gauged Wess-Zumino-Witten models \cite{GaKu,KaScho}, the
$G/H$ coset theories have a natural nilpotent BRST operator
\cite{KaScht} acting on the tensor product of highest weight modules of
the current algebra of $G$ and $H$, respectively.  One of the main
themes of these lectures will be to analyze this complex, and, in
particular to prove that its cohomology indeed yields the correct
spectrum of states for the coset theory.

The BRST formulation of coset theories is a fairly old problem. In the
case when $H$ is abelian, its cohomology was computed already in the
original paper \cite{KaScht}.  For a general $G/H$ theory,  another
complex was proposed in \cite{BMPnp}, which was based on the free field
realization of the  WZW model with group G.  Although it was possible
in this formulation to compute the cohomology explicitly, and thus
derive \eg branching functions for arbitrary coset theories, the
construction could not be recast in a covariant formulation from the
point of view of conformal field theory.  So a complete treatment of
the correlation functions for these models was not possible.  The
extreme case where the complete group is gauged away, $G/G$ coset
models, defines a topological field theory.  It was recently observed
in this simpler setting that a suitable choice of the module describing
the states of the $H$ (=$G$) sector of the theory allows just such a
covariant description, as well as an explicit computation of the BRST
cohomology \cite{Frprc,Sonna,HuYu,Sonnb,Sonnc}, in very much the same
manner as in $2d$ gravity.  One of the results we present here is a
general discussion of the BRST cohomology for an arbitrary $G/H$ coset
theory. We also construct the free field resolutions for these coset
models, and demonstrate how the corresponding extended BRST complex can
be reduced to our previous formulation \cite{BMPnp}.

As is noted several times above, the computations here appear
superficially similar to those in $2d$ gravity.  Indeed there exist
free field realizations of affine Kac-Moody algebras, and all
physically interesting irreducible highest weight modules admit
resolutions in terms of (twisted) Wakimoto modules
\cite{FFcmp,BeFe,BMPcmpo,BMPsb,FKW}, which take the role played by
Feigin-Fuchs modules in the case of the Virasoro algebra.  Thus, one
may first establish the BRST cohomology of a tensor product of two
(oppositely twisted) Wakimoto modules, and then project onto the
irreducible module using a suitable resolution.  However, it will
become apparent that the two contexts are really quite different, the
difference being due to a rather subtle property of Wakimoto modules;
namely, that they are essentially defined by their semi-infinite
cohomology with respect to a (twisted) maximal nilpotent
subalgebra of the affine Kac-Moody algebra \cite{FFcmp}.  We will show
that in fact a natural modification of standard results on
semi-infinite cohomology of Lie algebras \cite{FGZ,LZcmpo} allows a
straightforward computation of the BRST cohomology for general coset
models without ever resorting to an explicit realization of the
Wakimoto modules! This should be compared to the situation in $2d$
gravity, where an analogous cohomological characterization of
Feigin-Fuchs modules does not exist unless they are isomorphic with the
Verma module or its dual.  This forces one to be more resourceful, by
either trying to exploit the explicit structure of the BRST operator
\cite{BMPcmpt}, or by using the inherent $SO(2,\CC)$ symmetry of the
problem to reduce the computation so that it can be treated by standard
methods \cite{LZcmpt}.

The overall structure of these lectures is that we first survey some
techniques of homological algebra, and then show how to use them in the
computation of the BRST cohomology of coset models and $2d$ gravity.
After introducing the basic objects of semi-infinite cohomology in
Section 2, we review standard computational techniques in Section 3.
They all utilize one kind or other of spectral sequence.  The new
result here is a generalization of the reduction theorem
\cite{FGZ,LZcmpo}.  In Section 4 we discuss cohomological definitions
of Verma modules and Wakimoto modules of an affine Kac-Moody algebra,
and summarize the corresponding resolutions of irreducible highest
weight modules.  With this machinery in hand, Section 5 is devoted to
analysing the BRST formulation of the coset models. In Section 6 we
compare it with that of $2d$ gravity, and present a simplified
derivation of the discrete states in $c<1$ models.

P.B.\ would like to thank the organizers of the XXV Karpacz Winter
School for the opportunity to present these lectures, and the Aspen
Center for Physics for hospitality while they were written up.
K.P.\ would like to thank the Theory Division at CERN for hospitality
while some of the work was carried out. We would like to thank
E.\ Frenkel and A.\ Voronov for discussions.

\section{Definitions and conventions}
\setcounter{equation}{0}

{}For a Lie algebra $\mb g$ and a module $V$, the standard Lie algebra
cohomology of $\mb g$ with values in $V$ (see \eg \cite{Knapp,Fuchs})
is simply the cohomology of the operator \eqn{Lalgdif}{d=\sum_{A}c^A [
\pi(e_A)+\half\pigh(e_A)]\,,} in the complex $\cC(\mb g,V)$, where
\eqn{Lalgcom}{\cC(\mb g,V)=\bigoplus_{n\geq 0}\cC^n(\mb g,V)\,,\qquad
\cC^n(\mb g,V)= V\otimes {\textstyle\bigwedge}^n\mb g^*\,.} Here $e_A$,
$A=1,\cdots , \dim\mb g$, are the generators of $\mb g$, which act on
$V$ in the representation $\pi(\cdot)$, and on
${\textstyle\bigwedge}^n\mb g^*$,  the  $n$-th exterior power of the
(restricted) dual $\mb g^*$,  by $\pigh(\cdot)$, which is induced from
the coadjoint action of $\mb g$ on $\mb g^*$. The ghost operators
$c^A=c(e^{A*})$ can be identified with the basis $\{e^{A*}\}$  in $\mb
g^*$, dual to the generators $e_A$.  It is convenient to introduce
antighost  operators, $b_A=b(e_A)$, which are canonically conjugate to
$c^A$, \ie $\{b_A,c^B\}=\de_A^B$.  Then we may identify
 $\pigh(e_A) = -\sum_{B,C}c^Bb_C f_{AB}{}^C$, where $f_{AB}{}^C$ are
the structure constants of $\mb g$.  One may notice that together the
ghosts $c^A$ and antighosts $b_A$ span the Clifford algebra of $\mb
g\oplus \mb g^*$, with a canonical pairing $\langle
x,y^*\rangle=y^*(x)$. One can identify ${\textstyle\bigwedge}^n\mb g^*$
with the subspace of the ghost Fock space  with ghost number $n$. In
this Fock space all $c^A$ act as creation operators, and the vacuum, to
which we assign ghost number (order)  zero, is annihilated by all
$b_A$.  The differential $d$ is nilpotent by virtue of the Jacobi
identities. We will denote the resulting cohomology of order $n$ by
$H^n(\mb g,V)$.

There is an obvious difficulty  when applying this construction to
infinite dimensional algebras, as the differential (\ref{Lalgdif})
becomes a series  in which an infinite number of terms may act
nontrivially on a given states in the complex. To circumvent this
problem, Feigin \cite{Feigin} proposed  to replace the space of forms
${\textstyle\bigwedge}^\bull\mb g ^*$ by a suitably defined space of
semi-infinite forms ${\textstyle\bigwedge}^{\infty/2+\bull}\mb g ^*$.
As will become clear in a moment, his construction has its origin in
the physicists treatment of the infinite negative energy problem in the
quantization of fermion fields.

To define the semi-infinite cohomology we will restrict both the
possible algebras and the possible modules.  The first restriction is
to $\ZZ$-graded Lie algebras.  Any such $\mb g =\bigoplus_{i\in\ZZ}\mb
g _i $ can be decomposed as \eqn{decom}{ \mb g =\mb n_-\oplus\mb
t\oplus\mb n_+\,, } where \eqn{decar}{\mb n_-=\bigoplus_{i<0 }\mb g
_i\,,\quad \mb t=\mb g _0\,,\quad \mb n_-=\bigoplus_{i>0}\mb g _i\,.}
Corresponding to (\ref{decom}), the space of semi-infinite forms
${\textstyle\bigwedge}^{\infty/2+\bull}\mb g^*$ is defined as the ghost
Fock space $\cF_{gh}$, which is generated by the ghost and antighost
operators acting on a vacuum state $\om_0$  satisfying \begin{eqnarray}
b(x)\om_0&=&0\,, \quad {\rm for} \quad x\in \mb t\oplus \mb n_+\,,\\
c(y^*)\om_0&=&0 \,,\quad {\rm for} \quad  y^*\in \mb n_-^*= (\mb
t\oplus\mb n_+)^\perp\,.  \end{eqnarray} We assign to $\om_0$ the ghost
number equal zero, and will refer to it as a ``physical vacuum.'' One
should note that, unlike previously, there are states with both
positive as well as negative ghost numbers, and this  will lead to
two-sided complexes later on.

To make this discussion more concrete, let us assume $\mb g$ is one
of the following:

\begin{itemize}
\item a finite dimensional Lie algebra, $\bg$,
\item an  (untwisted) affine Kac-Moody algebra, $\widehat \bg$
($\widehat \bg=L\bg\oplus\CC k\oplus\CC d$
is  the centrally extended loop algebra of $\bg$),
\item the Virasoro algebra \Vir.
\end{itemize}

We will denote by $\De_+$ and $\De_-$ the space of positive roots and
negative roots of $\bg$, respectively, and by $\De = \De_+ \cup \De_-$
the total root space.  Similarly for $\widehat \bg$ we  have
$\widehat\De$, $\widehat\De_+$ and $\widehat\De_-$. We recall that in
terms of finite dimensional roots we can express the affine roots as
follows ($\widehat\De=\widehat\De_+\cup\widehat\De_-$) \eqn{affroots}{
\widehat\De_+=\{\al+n\de\,|\,\al\in\De_+,\,\,n\geq 0\}\cup
\{\al+n\de\,|\,\al\in\De_-,\,\,n> 0\}\cup \{n\de\,|\,n>0\} \, ,} where
$\de$ is the (imaginary) root dual to $d=-L_0$.  Further, we will
denote by $P$ (resp.\ $P_+$) the space of integral  weights
(resp.\ integral dominant weights) of $\bg$, and by $P^k$
(resp.\ $P_+^k$) the space of affine integral weights (affine
integrable weights) of $\widehat \bg$ at level $k$. We recall that an
affine weight $\widehat\la\in P^k$ can be represented by its finite
dimensional projection and the level $k$ as $\widehat\la=\la+k \La_0$,
where $\La_0$ is dual to $k$. For convenience we will usually use this
finite dimensional parametrization. We will also denote by $W$  the
Weyl group of $\bg$, and by $\widehat W$ that of $\widehat \bg$.  The
latter is isomorphic to the semi-direct product $\widehat W = W\times
T$, where $T$ is the long-root lattice, such that for $\widehat w=t_\ga
w$, $\widehat w\la=w\la+k\ga$ \cite{Kac}.

The $\ZZ$-grading for $\bg$ and $\widehat \bg$, which will be denoted
by $\deg(\cdot)$,  is defined by the height function in the  root
space. Explicitly, for a finite dimensional $\bg$  we have
\eqn{degree}{\deg(\mb t)=0\,,\quad \deg(\bg_\al)=\hgt(\al)=(\rh,\al)
\,,\quad \al\in \De\,,} where  $\rh$ is the element of $\mb t^*$ such
that $(\rh,\al_i^\vee)=1$ for the simple roots  $\al_i$,
$i=1,\ldots,\ell={\rm rank}\,\bg$. Then (\ref{decom}) is simply the
Cartan decomposition of $\bg$,
\eqn{XYZa}{\bg=\bn_-\oplus\bt\oplus\bn_+=(\bigoplus_{\al\in\De_-}
\bg_\al)\oplus\bt\oplus(\bigoplus_{\al\in\De_+}\bg_\al)\,.}

A similar decomposition holds for the affine Kac-Moody algebra
$\widehat \bg$, if we take $\widehat \rh=\rh+h^\vee\La_0$, where
$h^\vee$ is the dual Coxeter number of $\bg$.

In $\Vir$ the grading is defined by the eigenvalues of $-L_0$.  In
all three cases, the zero-degree subalgebra is abelian, and the space
of semi-infinite forms has a weight space decomposition with respect to
it. All of the above carries over to  subalgebras of  any of these
three algebras.

The second restriction is placed on the possible modules, we restrict
the class of $\mb g $-modules to the category $\cal O$ \cite{BGG,Kac}.
\thm{omodul}{Definition}{A module $V$ is in the category $\cal O$ if
\begin{enumerate} \item $V$ has a weight space decomposition
$V=\bigoplus_{\la\in P(V)}V_\la$, with finite dimensional weight spaces
$V_\la$.  \item There exists a set of weights $\la_1,\cdots,\la_n$ such
that  $P(V)\subset \bigcup_{i=1}^nD(\la_i)$, where $D(\la)$ denotes the
set of all descendant weights $\mu$ of $\la$, $D(\la)=\{\mu\in P\,|\,
\la-\mu\in\De_+\}$.  \end{enumerate}}

If $\mb g=\mb a$ is some subalgebra of the above three algebras ($\bg$,
$\wbg$ or $\Vir$), we will consider only such $\mb a$-modules $V$ which
are also $\mb t$-modules, with the weight space decomposition as
above.  A simple consequence of this definition is that $V$ is locally
$\mb n_+$-finite,\footnote{In fact, this sole property may be used to
restrict the modules in a more general approach to semi-infinite
cohomology \cite{Vor}.} \ie for any $v\in V$ the subspace $\cU(\mb
n_+)v$ is finite dimensional, where $\cU(\mb n_+)$ is the enveloping
algebra of $\mb n_+$. This category $\cal O$ of $\mb g $ modules is
rather large -- it includes Verma modules, Feigin-Fuchs and Wakimoto
modules, their tensor products, submodules, quotients, duals, etc.

{}For a module $V$ in the category $\cal O$ we set
\eqn{SScom}{\cC^{\infty/2+\bull}(\mb g ,V)=V\otimes
{\textstyle\bigwedge}^{\infty/2+\bull}\mb g ^*\,,} and consider an
operator $d:\cC^{\infty/2+n}(\mb g ,V)\rightarrow
\cC^{\infty/2+n+1}(\mb g ,V)$, \eqn{SSdiff}{ d=\sum_A c^A\pi(e_A)
-\half:\sum_{A,B,C}c^Ac^Bb_C f_{AB}{}^C : + c(\be)\,,} where $:\,\,:$
denotes the normal ordering with respect to $\om_0$, and $\be$ is some
constant element of $\mb g ^*$.  Since only a finite number of terms
contribute to the action of $d$ on any given state, this operator is
certainly well defined.  The following condition for the nilpotency of
$d$ is a classic result (see \eg \cite{Feigin,FGZ})
\thm{nilp}{Theorem}{If the total central charge of the representation
\eqn{gen}{\Pi(e_A)=\{d,b_A\}=\pi(e_A)-:\sum_{B,C}c^Bb_C f_{AB}{}^C
:\,,} of $\mb g $  ($\mb g=\bg$, $\wbg$ or $\Vir$) in
$\cC^{\infty/2+\bull} (\mb g ,V)$
 vanishes, then there exists a $\be $ such that $d$ defined in
(\ref{SSdiff}) is nilpotent, \ie $d^2=0$.}

In fact, for the algebras considered in these lectures, we may always
set $\be$ to zero by modifying the normal ordering prescription, \eg by
ordering with respect to the $SL(2,\RR)$ invariant vacuum of conformal
field theory.  Also, we note that the spaces $\cC^{\infty/2+n}(\mb g
,V)$ with the representation defined in (\ref{gen}) are in the category
$\cal O$.

We will denote the corresponding  semi-infinite cohomology classes of
ghost number $n$ by $H^{\infty/2+n}(\mb g ,V)$, or, sometimes, by
$H^{\infty/2+n}(d,V)$.

Finally, given a subalgebra $\mb h\subset \mb g $ one defines a complex
$\cC^{\infty/2+\bull}(\mb g ,\mb h;V)$ of semi-infinite cohomology
relative to $\mb h$, as the subcomplex which consists of those chains
in $\cC^{\infty/2+\bull}(\mb g ,V)$ which are annihilated by $b(x)$ and
$\Pi(x)$ for all $x\in\mb h$.

This concludes our review of basic facts about the semi-infinite
cohomology. For more detailed information and proofs of the results
cited above the reader should consult one of the basic papers
\cite{Feigin,FGZ,LZcmpo}, and \cite{Vor} for an abstract definition of
semi-infinite cohomology as a derived functor of semi-invariants.

\section{Basic computational techniques}
\setcounter{equation}{0}

As we will see in the following sections, physically relevant complexes
typically come equipped with some additional structure.  This is
usually sufficient to allow an explicit computation of the cohomology
using spectral sequences, a standard technique from homological algebra
(see \eg \cite{BT,HS}).  In this section we will review such methods,
keeping in mind their applications later on. Our discussion will
proceed from the most general spectral sequences to the more
specialized ones, which arise in the computation of Lie algebra
cohomology. In the latter case most of the results will also be valid
for the semi-infinite cohomology, and, unless some subtleties are
present, we will simplify the notation and write $n$ instead of
$\infty/2+n$.

\subsection{Cohomology of a filtered (graded)  complex}

\label{generalspsq}
\newcommand{\lC}[1]{\cC^{#1}}
Consider a complex $(\cC,d)$ of complex vector spaces, where
$\cC=\bigoplus_n\cC^{n}$ and the differential $d:\lC{n} \to\lC{n+1}$.
Suppose that there is an additional gradation, such that for each order
(ghost number) $n$,\footnote{This is stronger than the usual assumption
that $\cC$ must be a filtered complex. A standard filtration in our
case, for which $\cC$ is isomorphic with the graded object, is given by
subspaces $F_p\cC^{n}=\bigcup_{k\geq p} \cC_k^{n}$.}
\eqn{SSaa}{\lC{n}=\bigoplus_{k\in \ZZ} \lC{n}_k \,.} We will refer to
the integer $k$ as the degree, and denote by $\pi_k$ the projection
onto the subspace of degree $k$. This gradation by the degree must
satisfy the following properties:\smallskip

\begin{enumerate} \item The differential $d$ has only terms of
nonnegative degree, \ie \eqn{SSab}{d=d_0+d_1+\ldots=d_0+d_>\,,} where
\eqn{SSac}{d_i:\lC{n}_k\to\lC{n+1}_{k+i}\,.} \item In each order only a
finite number of nontrivial degrees are present, \ie for each $n$,
spaces $\lC{n}_k$ are nontrivial for a finite number of $k$'s.
\end{enumerate}

The spectral sequence associated with such a gradation allows a
systematic computation of the cohomology classes $H(d,\cC)$.  It is a
sequence of complexes $(E_r,\de_r)_{r=0}^\infty$, such that
\eqn{SSdef}{E_0=\cC\,,\quad E_1=H(d_0,\cC)\,,\quad
E_{r+1}=H(\de_r,E_r)\,,\quad r=1,2,\ldots\,.}  The differentials are
defined recursively, beginning for $r=0$ with the definition
$\de_0=d_0$.  Now, for $r\geq1$, notice that $\psi_k\in \cC$ represents
an element in $E_r$ (possibly a trivial one) if and only if there exist
$\ps_{k+1},\ldots,\ps_{k+r-1}$ such that \eqn{SSrter}{\pi_i
d(\psi_k+\ps_{k+1}+\ldots+\ps_{k+r-1})=0\,,\qquad {\rm for} \quad
i=k,\ldots,k+r-1\,.} Then we may define $\de_r:E_r\to E_r$ via
\eqn{SSdifr}{\de_r\psi_k=\pi_{k+r}
d(\psi_k+\ps_{k+1}+\ldots+\ps_{k+r-1})\,.}

The $E_1$ term of this sequence is obviously well defined, because
(\ref{SSab}) together with $d^2=0$ imply that $d_0^2=0$. To verify that
the subsequent terms are also well defined -- in particular, that
$\de_r$ are nilpotent operators in $E_r$ -- becomes more and more
tedious, so one usually resorts to more abstract techniques
\cite{BT,HS} rather than using the explicit formulae (\ref{SSrter}) and
(\ref{SSdifr}).  Nevertheless, it is quite illuminating to work out by
hand at least the next two terms (see, \eg \cite{BMPsrg}).  It then
becomes clear that the elements of $E_r$ are those cohomology classes
of $d_0$ which can be extended to approximate cohomology classes of $d$
through $r$ degrees.

A spectral sequence $(E_r,\de_r)$ becomes a useful device provided it
converges, which means that the spaces $E_r$ stabilize, \ie
\eqn{SSstab}{E_r=E_{r+1}=\ldots =E_\infty\,,} for some $r\geq 1$.
Obviously, this requires
\eqn{SSdst}{\de_r=\de_{r+1}=\ldots=\de_\infty=0\,.} In such a case one
also says that the sequence collapses at $E_r$.

As might be expected, one can prove the following fundamental theorem
(see, \eg \cite{BT,HS}).

\thm{SSlimit}{Theorem}{For a graded complex $(\cC,d)$ as above
\eqn{SSlim}{E_\infty\cong  H(d,\cC)\,.}}

Most of spectral sequences, which we will encounter later in these
lectures, collapse at the first term, because $E_1$ turns out to be
nontrivial only in either one ghost number or one degree (or both).  As
the differentials $\de_r$, $r\geq 1$, increase both the order and the
degree, they must vanish, which implies $E_1\cong  E_\infty$, or,
equivalently, $H(d_0,\cC)\cong  H(d,\cC)$.  Clearly this happy
circumstance will often be the result of a clever choice for the
definition of degree, which splits the calculation in this opportune
way.

\subsection{Spectral sequences of a double complex}

With a double complex $(\cC,d,d')$ \eqn{SSdbl}{d:\cC^{p,q}\to
\cC^{p+1,q}\,,\quad d':\cC^{p,q}\to \cC^{p,q+1}\,,}
\eqn{SSdbd}{d^2=d'{}^2=dd'+d' d=0\,,} one can associate two spectral
sequences, $(E_r,\de_r)$ and $(E_r',\de_r')$, arising from the grading
of the underlying single complex $(\cC,D)$,
\eqn{SSdbsi}{\cC^{n}=\bigoplus_{p+q=n}\cC^{p,q}, \quad D=d+d'\,,}
 by $p$ and $q$, respectively. In the first case we have
$\cC_p=\oplus_q\cC^{p,q}$, which gives \eqn{SSdbf}{E_1\cong
H(d',\cC)\,\quad \de_1=d\,,} while in the second case
$\cC_q'=\oplus_p\cC^{p,q}$, and \eqn{SSdbdsf}{E_1'\cong
H(d,\cC)\,\quad \de'_1=d'\,.}

One should note that the spaces $E_r$ and $E'_r$ in these spectral
sequences are doubly graded, and the action of the induced
differentials $\de_r$ and $\de_r'$ with respect to this gradation is
\begin{eqnarray} \de_r&:&E_r^{p,q}\longrightarrow E_r^{p+r,q-r+1}\,,\\
\de_r'&:&E_r'{}^{p,q}\longrightarrow E_r'{}^{p-r+1,q+r}\,.
\end{eqnarray}

In both cases the limit of the spectral sequence $E_\infty$ or
$E'_\infty$, if it exists, yields the cohomology of the complex $\cC$
with the differential $D=d+d'$. The following frequently used theorem
summarizes in which sense the order of computing the cohomologies of
$d$ and $d'$ can be interchanged.  \setcounter{SSdbex}{1}
\thm{SSdbex}{Theorem}{Suppose
 that in a double complex $(\cC,d,d')$ both spectral sequences
$(E_r,\de_r)$ and $(E_r',\de_r')$ collapse at the second term. Then
\eqn{SScol}{\bigoplus_{p+q=n}H^{p}(d,H^{q}(d',\cC)) \cong
\bigoplus_{p+q=n}H^{q}(d',H^{p}(d,\cC))\,.}}

\proof \cite{BT} We simply have \begin{eqnarray} \label{SSpf}
\bigoplus_{p+q=n}H^{p}(d,H^{q}(d',\cC)) &=&\bigoplus_{p+q=n}E_2^{p,q}
\nonumber \\ &=&H^{n}(D,\cC) \nonumber \\
&=&\bigoplus_{p+q=n}E_2'{}^{p,q}=\bigoplus_{p+q=n}H^{q}(d',H^{p}(d,\cC))\,.
\end{eqnarray}

\subsection{``Split and flip'' spectral sequences}

Throughout this subsection we assume that $\mb g$ is either  $\bg$, or
$\widehat \bg$ or  $\Vir$ or a subalgebra of those, and that all
modules are highest weight modules in the category $\cal O$. We recall
that in particular this means that  any module  $V$ is also a module of
$\bt$, $\widehat \bt$ or $\Vir_0$, and there is a single state $v_\La$,
with the highest weight $\La$, and the weight space decomposition of
$V$ is $V=\bigoplus_{\la\in P(V)}V_\la$, where $P(V)\subset D(\La)$.

 Of particular interest is the specific case in which $V$ is the tensor
product of two highest weight modules $V_1$ and $V_2$, with highest
weights $\La_1$ and $\La_2$, respectively.  We show here how to
construct a spectral sequence that allows an estimate of the relative
cohomology of the tensor product module $V_1\otimes V_2$, in terms of
the  cohomologies of $V_1$ and $V_2$.  This will be achieved by
introducing a family of degrees on the complex $\cC(\mb g,V_1\otimes
V_2)$, which are a natural generalization of the $f$-degree defined in
\cite{FGZ,LZcmpo}.

Let $f$ be an arbitrary integer valued function on the root lattice,
such that\footnote{Here, and in the remainder of this section, we use
the same notation for the roots and weights of all three algebras.}
\eqn{XYZb}{f(\al_i)\not=0\,,\quad
f(n_1\al_i+n_2\al_j)=n_1f(\al_i)+n_2f(\al_j)\,,} for all simple roots
$\al_i$, $\al_j$, and integers $n_1$ and $n_2$.  Examples of such
functions, which we will consider in the next section, are obtained  by
taking (for $\mb g = \bg$)
\eqn{XYZc}{f(\al)=(\rh,w\al)\,,\quad \al\in\De\,, \,\, w\in W\,,} or
similarly (for $\mb g = \widehat \bg$)
\eqn{XYZd}{f(\al)=(\widehat \rh,w\al)\,,\quad \al\in \widehat \De\,,
\,\, w\in \widehat W\,.} We will denote such $f$'s by $f_w$.

Obviously, using $f$,  we can decompose $\mb g$ into a direct sum
\eqn{gfdec}{\mb g=\mb n_-^f\oplus \mb t\oplus \mb n_+^f\,,} where $\mb
n_+^f$ and $\mb n^f_-$ are subalgebras corresponding to the positive
and negative roots defined with respect to $f$. We may also extend $f$
to a highest weight module $V$ (with highest weight $\La$) by setting
\eqn{fonv}{f(v)=f(\la-\La)\,, \quad {\rm for} \quad v\in V_\la\,.}

Similarly, we can use the weight space decomposition to extend $f$ to
the ghost Fock space,  $\cF_{gh}$, by setting $f(\om_0)=0$. Note that
in this case the ghosts and anti-ghosts corresponding to $\mb t$ will not
change the value of $f$.

Let us concentrate on the relative cohomology of $\mb g$ in $V_1\otimes
V_2$.  One of the effects of passing to the relative cohomology is that
we drop the ghosts and anti-ghosts of $\mb t$, \ie $\cF_{gh,rel} \cong
{\textstyle\bigwedge}(\mb n_-^f\oplus \mb n_+^f)^*$. This space
decomposes into a tensor product $\cF_{gh,rel}=\cF_{gh,-}^f\otimes
\cF_{gh,+}^f\cong {\textstyle\bigwedge}(\mb n_-^f)^*\otimes
{\textstyle\bigwedge}(\mb n_+^f)^*$.  Combining that with the tensor
product structure of the module itself, we are led to consider the
following decomposition of the spaces in the complex
 \eqn{crtpp}{\cC^n(\mb g,\mb t;V_1\otimes
V_2)=\bigoplus_{p+q=n}\bigoplus_{\la} \,\cC^{p}_\la(\mb
n_-^f,V_1)\otimes \cC^{q}_{-\la}(\mb n_+^f,V_2)\,,} where
$\cC_\la(\cdot,\cdot)$ denotes the subspace with the weight $\la$.
Note that because $\cC^n(\cdot,\cdot)$ is a module in category $\cal
O$, the sum in (\ref{crtpp}) is in fact finite both with respect to $n$
and $\la$. (This is clear  for $\mb g =\bg$, and for  $\widehat \bg$
and $\Vir$ becomes obvious if we remember that all ghost and anti-ghost
creation operators have positive energy, which is a part of the affine
weight.) We emphasize that  at this point (\ref{crtpp}) is merely an
equality between {\it vector spaces}, and that the assignment of $\mb
n_-^f$ to $V_1$ and $\mb n_+^f$ to $V_2$ is completely arbitrary.

However, it is easy to see that the differential $ d$ in $\cC(\mb g,\mb
t;V_1\otimes V_2)$  is of the form $d=d_-+d_++\cdots$, where $d_-$ and
$d_+$ are the differentials in $\cC_\la(\mb n_-^f,V_1)$ and
$\cC_{-\la}(\mb n_+^f,V_2)$, respectively, while the additional terms
correspond to the action of $\mb n_-^f$ and $\mb n_+^f$ on $V_2$ and
$V_1$, respectively, and 3-ghost terms that arise when $\mb n_-^f$ and
$\mb n_+^f$ do not commute.  This form of the differential  suggests
the introduction of a degree so that, at the first term of the spectral
sequence, (\ref{crtpp}) becomes an equality between {\it complexes}.
Such a degree, which we denote $f$deg, may be defined via
\eqn{filt}{f{\rm deg}((v_1\otimes\om_1)\otimes( v_2\otimes
\om_2))=f(v_1)+f(\om_1)-f(v_2)-f(\om_2)\,,} where $v_1\in V_1$, $v_2\in
V_2$, $\om_1\in \cF_{gh,-}^f$ and $\om_2\in\cF_{gh,+}^f$.  Note that on
the ghosts, \eqn{fongh}{f{\rm deg} (\prod_\al c^\al\prod_\be
b_\be\,\om_0)= \sum_\al|f(\al)|-\sum_\be|f(\be)|\,.} In other words
each ghost $c^\al$ increases $f$ by $|f(\al)|$, while $b_\be$ decreases
$f$ by $|f(\be)|$.  By virtue of the triangle inequality satisfied by
$|f(\cdot)|$ applied to the 3-ghost terms, and recalling that  $\Pi(\mb
t)\equiv 0$ on the subcomplex
 of relative cohomology, it is easy to check that the $f{\rm deg} = 0$
term of the differential in the complex on the l.h.s.\ is indeed the
sum of the differentials in complexes on the r.h.s.\ of (\ref{crtpp}),
\ie $d_0=d_++d_-$!  As we have argued above, this filtration is finite,
and so the corresponding spectral sequence converges. Thus we have
shown the following theorem.  \setcounter{flip}{2} \thm{flip}{Theorem}{
There exists a spectral sequence $(E_r,\de_r)$ such that
\begin{eqnarray} E_1^n&\cong & \bigoplus_{p+q=n}\bigoplus_\la
H^p_\la(\mb n_-^f,V_1)\otimes H^q_{-\la}(\mb n_+^f,V_2)
 \,,\label{collap}\\ E_\infty^n &\cong & H^n(\mb g,\mb t;V_1\otimes
V_2)\,.  \end{eqnarray}}

For obvious reason we will refer to this sequence as the ``split
and flip'' spectral sequence.

\subsection{Reduction theorems}

Let us now consider the situation when the cohomology of one of the
complexes on the r.h.s.\ of  (\ref{crtpp}) is particularly simple;
namely, it has only one state, at  ghost number number zero, \eg
\eqn{redas}{ H^q(\mb n_+^f,V_2)= \de^{q,0}\,\CC_{-\la}\,.} Then, for
each $n$, the sum in (\ref{collap}) collapses to just one term with
$p=n$.  In fact, as we will now show, the entire spectral sequence
collapses, and we obtain the following reduction theorem.

\setcounter{redthm}{3} \thm{redthm}{Theorem} {For a  module $V_2$
satisfying (\ref{redas}) the spectral sequence of Theorem \ref{flip}
collapses at $E_1$ and we obtain \eqn{redcoh}{H^n(\mb g,\mb
t;V_1\otimes V_2)\cong H^n_\la (\mb n_-^f,V_1)\otimes \CC_{-\la}\,.}}
\def\fdegg{f{\rm deg}} \proof We observe that in the case of
 relative cohomology, there is an additional relation between the
$f{\rm deg}$ of the two factors in (\ref{filt}), namely
\eqn{JMa}{\big(f(v_1) + f(\om_1)\big) + \big(f(v_2) + f(\om_2)\big) = -
f(\La_1+\La_2)={\rm const}\,.} This shows that $E_1^n$ can be
nontrivial in only one $f$-degree -- the same one for all $n$. Thus the
sequence must collapse, as we have discussed at the end of Section
\ref{generalspsq}.$\Box$

For $f=-f_{w=1}=-{\rm deg}$ this theorem becomes the reduction theorem
of \cite{LZcmpo}.

\subsection{Resolutions}

A direct calculation of the cohomology on a given module will usually be
complicated, and yet one can often find special modules for which
the calculation is immediate.  For an arbitrary $\mb g$-module, $V$,
a promising -- and usually necessary -- line of attack is
find a description of $V$ in terms of these special modules.
In fact, the possibility of carrying out such a procedure is a good test
as to whether one is working in a sensible
category of modules!  A decomposition of $V$ into its ``building
blocks'' is achieved in terms of a resolution.

\def\cR{{\cal R}}
\thm{resolution}{Definition}{ We say that a complex $(\cR,\de)$ of
$\mb g$-modules $\cR^n$ and $\de:\cR^n\to\cR^{n+1}$ intertwining with the
action of $\mb g$ is a resolution of the $\mb g$-module $V$ if
\eqn{rescoh}{H^n(\de,\cR)\cong  \de^{n,0}V\,.}}

Resolutions are said to be one-sided if we can set $n\geq0$,
or two-sided if $n\in\ZZ$, finite or infinite, etc.

A typical application is as follows. Suppose we want to compute $H(\mb
g,V)$, and we know how to write down a resolution of $V$ in which
$H(\mb g,\cR^n)$ are computable and simple. By replacing $V$ with its
resolution we have \eqn{replace}{H^n(\mb g,V)=H^n(\mb g,H^0(\de,{\cal
R}))\,.} Clearly the  form of the r.h.s.\ suggests that we consider the
double complex $(\cC,\de,d)$, $\cC^{n,m}=\cR^n\otimes
{\textstyle\bigwedge}^m\mb g^*$, and see whether Theorem \ref{SSdbex}
can be applied to change the order of cohomologies in (\ref{replace}).
If so, then we may exploit the simple structure of $H(\mb g,\cR^n)$.
Indeed, in all the cases we discuss later on this is the best way to
proceed.

\section{Resolutions of highest weight modules of  affine Kac-Moody
algebras}
\setcounter{equation}{0}

In this section we review  two classes of resolutions of irreducible
highest weight modules of affine (untwisted) Kac-Moody algebra $\wbg$:
a one-sided resolution in terms of Verma modules, and a set of two
sided resolutions in terms of Wakimoto modules.  A similar
constructions for the Virasoro algebra will be briefly summarized in
Section 6.

\subsection{Verma modules and the BGG-resolution}

Recall that a Verma module $M_{\La}$ of $\wbg$, with highest weight
$\La$, is freely generated by $\wbn_-$ from a highest weight state
$v_{\La}$ such that $nv_\La=0$,  $n\in\wbn_+$ and $tv_\La=\La(t)v_\La$,
$t\in\widehat \bt$.  We will denote by $M_\La^\star$ the contragradient
Verma module \cite{Diximier}, which is dual to $M_\La$ with respect to
the canonical pairing in $\cU(\wbg)$, and co-free with respect to
$\wbn_+$. It turns out that both of these modules can also be defined
in a rather surprising manner \cite{BGG,Feigin}.

\thm{cohver}{Theorem}{ A Verma module $M_{\La}$ of $\wbg$ is completely
characterized by the following two conditions \begin{itemize} \item
$M_{\La}$ is a module in category $\cal O$.  \item As a $\widehat
\bt$-module, \eqn{vercoh}{H^{\infty/2+n}(\wbn_-,M_{\La})\cong
\de^{n,0}\,\CC{}_{\La}\,.} \end{itemize}} By Poincar\'e duality in
semi-infinite cohomology \cite{FGZ,LZcmpo}, a similar theorem holds for
the contragradient Verma module $M_{\La}^\star$, except that one must
replace $\wbn_-$ with $\wbn_+$.

\proof We sketch the proof \cite{BGG}.  To calculate (\ref{vercoh}),
introduce an (increasing) filtration on the complex ${\cal
C}^{\infty/2+\bullet} (\widehat{\bf n}_-,M_\La)$ defined, on the basis
of monomials, by \eqn{filters}{F^p{\cal C} = \{ v=e_{-\al_1}\cdots
e_{-\al_r}b_{-\be_1}\cdots b_{-\be_s}v_\La\, |\, r+s\leq p\}\,.}  Using
the fact that $M_\La$ is freely generated by $\widehat{\bf n}_-$ one
finds that the first term in the associated spectral sequence is the
so-called Koszul complex \cite{Knapp,HS}, for which it is easy to show
that the cohomology is one-dimensional and generated by $v_\La$. The
spectral sequence then collapses to give (\ref{vercoh}). Note that one
uses here only $\wbn_-\otimes\widehat\bt$-module structure of $M_\La$!

As for the  opposite part of the theorem, let $V$ be a module in the
category $\cal O$ whose cohomology is the same as that of $M_\La$.
Since  in (\ref{cohver}) $n\leq0$, we have $H^{\infty/2+0}(\wbn_-,V)
\cong V/\wbn_-V$. (Note that such a characterization would not be valid
if $n>0$  terms were present in the complex!) Thus the module $V$ is
generated by $\cU(\wbn_-)$ acting on the higest weight state $v'_\La$,
corresponding to the nontrivial 0-th cohomology. Equivalently, we have
a surjection  $\si$ from the Verma module $M_\La$ onto $V$, defined by
$\si(v_\La)=v_\La'$. This gives a short exact sequence $0\rightarrow
K\rightarrow M_\La \rightarrow V \rightarrow 0$, where $K=\ker\,\si$.
{}From the corresponding long exact sequence in cohomology  \cite{HS,BT}
we find that the cohomology of $K$ must be trivial.  This implies that
$K=0$,  and thus $V\cong M_\La$,  as otherwise we would have a
nontrivial module in the category $\cal O$ with zero cohomology, which
is clearly impossible since it contains ``highest'' weights.$\Box$

Before we formulate the next theorem, on the construction of
resolutions in terms of Verma modules, we recall that for a given
$w'\in\widehat W$, the length of $w'$ is defined as $\ell(w')=
|\Ph(\ w')|$, in terms of the set of roots $\Ph(w')=\widehat \De_+\cap
w'(\widehat \De_-)$ \cite{Kac}.  We also introduce a twisted action,
denoted by $*$, of $\widehat W$ on the weights,
$w'*\La=\overline{w}{}'(\La+\rh)-\rh +(k+h^\vee)\ga$ for $w'= t_\ga
\overline w{}'\in\widehat W$, $\La\in P^k$.  \setcounter{bla}{1}
\thm{bla}{Theorem}{ Let $\La\in P_+^k$
 be an integrable weight of $\wbg$. Then there exists a resolution of
the irreducible module $L_\La$ in terms of Verma modules given by the
complex $({\cal M}^{(n)}_\La,\de^{(n)})$, $n\leq 0$, such that
\eqn{XYZe}{{\cal M}^{(n)}_\La=\bigoplus_{\{w'\in \widehat W\,|\,
\ell(w')=-n\,\}} M_{w'\star\La}\,,} and the differentials $\de^{(n)}$
are determined, up to  combinatorial factors, by the singular vectors
in $M_\La$.}

\proof See \cite{BGG} for the proof for finite dimensional Lie algebras
and  \cite{GL,RoWa} for its generalization to  Kac-Moody algebras.

\subsection{Wakimoto modules and free field resolutions}

It is natural to ask whether any other $\wbg$-modules can be
characterized in a way similar to the cohomological characterization of
Verma modules and contragradient Verma modules, using subgroups
isomorphic with  $\wbn_-$ and/or $\wbn_+$. This question motivated the
following construction of Wakimoto modules \cite{Wak} proposed by
Feigin and Frenkel some three years ago \cite{FFcmp}.

For a given $w\in W$, consider a subgroup  $\wbn{}^w_+=w_\infty\cdot
\wbn_+\equiv w_\infty\wbn_+w_\infty^{-1}$, where, formally,$w_\infty=
\lim_{N\rightarrow\infty}w_N $, $w_N=w t_{N\rh}\in \widehat W$.  The
action of this infinite twist should be  understood as
\begin{equation} \wbn{}_+^w = \{x\in \wbg\,|\, \exists_{N_0}
\forall_{N\geq N_0}\, x\in w_N\cdot\wbn_+\}\,.  \end{equation}
Explicitly,
\eqn{XYZf}{\wbn{}^w_+=\bigoplus_{\widehat\al\in\widehat\De^w_+}
\wbg_{\widehat\al}} where
\eqn{XYZg}{\widehat\De_+^w=\{\widehat\al=n\de+w\al\,|\,\al\in\De_+\,,
n\in\ZZ\} \cup\{\widehat\al=n\de\,|\, n>0\}\,,} \ie $\wbn{}_+^w$ is a
sum of the current algebra based on $\bn_+^w=w\bn_+ w^{-1}$ (a twisted
nilpotent subalgebra of $\bg$), and the positive modes of  currents
corresponding to the Cartan subalgebra $\bt$.  An equivalent way of
characterizing $\widehat\De_+^w$ is as those roots for which (see
Section 3.3) $f_{w_N}(\widehat\al)=(\widehat\rh,w_N\widehat\al)$ is
positive for $N$ sufficiently large.

Let us also introduce \eqn{rootsdec}{ \widehat \De_+^{w,(\pm)}=\widehat
\De_+^w\cap\widehat\De_\pm\,,\quad \widehat \De_-^{w,(\pm)}=
(-\widehat\De_+^w)\cap\widehat\De_\pm\,,} and denote the corresponding
subgroups by $\wbn_+^{w,(\pm)}$ and $\wbn_-^{w,(\pm)}$, respectively.
We then have decompositions
$\wbg=\wbn{}_-^w\oplus\widehat\bt\oplus\wbn{}_+^w$ and $\wbn^w_\pm=
\wbn_\pm^{w,(-)}\oplus \wbn_\pm^{w,(+)}$. Also, in analogy with the
usual length $\ell$ above, one can introduce a ``twisted length''
 $\ell_w$ on $\widehat W$, defined by \cite{FFcmp,BMPcmpo}
\eqn{twistedlen}{\ell_w(w')=|\Ph^{w,(+)}(w')|-|\Ph^{w,(-)}(w')|\,,\quad
\Ph^{w,(\pm)}(w')=\widehat\De_+^{w,(\pm)}\cap w'(\widehat\De_-)\,.}

With this somewhat elaborate machinery we have \cite{FFcmp}
\thm{wakmod}{Definition}{For $w\in W$, a  Wakimoto module
$F^w_\La$ of an affine Kac-Moody algebra $\wbg$ is a module such that
\begin{itemize} \item $F^w_\La$ is a $\wbg$-module in category $\cal
O$. \item As a $\widehat \bt$-module \eqn{wakscohom}{
H^{\infty/2+n}(\wbn{}_+^w,F^w_{\La})\cong  \de^{n,0}\,\CC{}_{\La}\,.}
\end{itemize}}

To show that this definition is not vacuous, one must construct
examples of such modules. In conformal field theory one can give an
explicit realization of Wakimoto modules as Fock spaces of a set of
conjugate first order bosonic free fields of conformal dimension
$(1,0)$ (one such pair for every root $\al \in \De_+^w$), and a set of
free scalar fields with background charge (as many as the rank of
$\bg$) (see \eg \cite{BMPrev} and the references therein).  The most
frequently used realization appears to be for $w=1$, but, as we will
see in the next section, the ability to implement the general twist
plays a crucial role in the construction of a resolution for the coset
module.

Although  free field realizations of the Wakimoto modules have been
discussed at some length in the literature, their cohomology has only
been considered in \cite{FFcmp}.  Let us briefly outline
here a proof of (\ref{wakscohom}).

As with Verma modules, the Wakimoto modules $F^w_\La$ constructed thus
far are particularly simple when viewed as $\wbn{}_+^w\oplus
\widehat\bt$ modules, namely, \eqn{explwakmto}{ F^w_\La \cong
\cU(\wbn_+^{w,(-)})\otimes \cU(\wbn_-^{w,(-)})^\star\otimes\CC_\La \,.}
We identify here $\cU(\wbn_+^{w,(-)})$ with the Verma module of
$\wbn_+^w$ built on the vacuum annihilated by the generators in
$\wbn_+^{w,(+)}$. Similarly we define the contragradient module of
$\wbn_+^w$ in the second factor. Finally, $\wbn_+^w$ acts trivially on
the third factor, which is introduced to shift the highest weight to
the desired value. In this explicit realization the cohomology
computation is almost trivial. Consider an $f$deg as in Section 3.3
with $f=f_\infty=\lim_{N\rightarrow\infty}f_{w_N}$.  Identifying
$\cU(\wbn_+^{w,(-)})$ with $V_1$ and $\cU(\wbn_-^{w,(-)})^\star$ with
$V_2$, the first term of the resulting ``split and flip'' spectral
sequence is given by \eqn{waksplit}{ E_1\cong\bigoplus_{n\leq
0}\bigoplus_{m\geq 0}
H^{\infty/2+n}\big(\wbn_+^{w,(-)},\cU(\wbn_+^{w,(-)})\big)\otimes
H^{\infty/2+m}\big(\wbn_+^{w,(+)},\cU(\wbn_-^{w,(-)})^\star \big)\,,}
and, by virtue of Theorem \ref{cohver}, it is nontrivial only for
$m=n=0$. Thus the sequence collapses, and we obtain (\ref{wakscohom}).

There remains the intriguing mathematical question as to whether
Wakimoto modules are uniquely determined by their cohomology. It is not
too difficult to show that the cohomology (\ref{wakscohom}) forces the
module to be free over $\wbn_+^{w,(-)}$, and co-free over
$\wbn_+^{w,(+)}$, which of course is manifest in the explicit
realization (\ref{explwakmto}).  To prove this one can consider two
Hochschild-Serre spectral sequences\footnote{We thank E.\ Frenkel for
suggesting this line of reasoning.} \cite{HoSe} corresponding to two
subgroups of $\wbn_+^w$.  Specifically, one easily sees that the
differential $d$ in the complex $\cC(\wbn_+^w,F)$ can be split into a
sum of two anticommuting differentials, $d=d_++d_-$, corresponding to
$\bn_+^{w,(+)}$ and  $\bn_+^{w,(-)}$.  Thus we have a structure of a
double complex. The two Hochschild-Serre spectral sequences are then
simply those arising via the usual bi-grading of this double complex,
as discussed in Section 3.2.  After some rather subtle analysis, the
freeness and co-freeness essentially follow from Theorem \ref{cohver}.
However, we were not able to complete the argument for the uniqueness
of the module, and we are not aware of any existing explicit proof of
this fact.

As expected we have the following resolutions \cite{FFcmp,BMPcmpo}.
\setcounter{brrrr}{2} \thm{brrrr}{Theorem}{For any  $\La\in P_+^k$ and
$w\in W$, there exists a resolution of $L_\La$ in terms of Wakimoto
modules given by a complex $(\cF^{w,(n)}_\La,\de^{w,(n)})$, where
\eqn{fockresolution}{{\cal F}^{w,(n)}_\La= \bigoplus_{\{ w'\in \widehat
W\,;\, \ell_w( w')=n\,\}} F^w_{w'*\La}\, ,\quad n\in\ZZ.}}

\def\wW{\widehat W} \def\Lap{\La^{(+)}} \def\Lam{\La^{(-)}}

In many applications in conformal field theory and topological field
theories considering only integrable weights is not sufficient.
Rather, we have to consider fractional levels, say \eqn{PBCm}{ k+\dCo =
{p\over p'}\,,\quad{\rm gcd}(p,p')=1\,,\quad {\rm gcd}(p',r^\vee)
=1\,,\quad p\geq\dCo\,,\quad p'\geq{\rm h}\,, } and weights of the form
\eqn{PBCn}{ \La = \Lap - (k+\dCo)\Lam\,, } with $\Lap\in P_+^{p-\dCo}$
and $\Lam\in P_+^{\vee\,p'-{\rm h}}$.  Here, $r^\vee$ is the ``dual
tier number'' of $\wbg$. We recall that in the simply laced case ${\rm
h}=\dCo\,,P_+=P_+^\vee$ and $r^\vee=1$.  Weights of the form
(\ref{PBCn}) are a subset of the class of so-called admissible weights
\cite{KWa}.

It turns out that the entire discussion of the case of
integrable representations can easily be extended to this more general
class of admissible weights. One can construct a complex as above in
which
\eqn{PBCr}{ \cF_\La^{w,(n)} = \bigoplus_{ \{ w'\in\wW | \ell_w(w') =
n\} }\ \cF_{ w'*\Lap - (k+\dCo)\Lam }\,,} and the differentials are
precisely the same as in the resolution for the integrable weight
$\La^{(+)}$ \cite{BMPsb,FKW}.
 We expect that this complex provides a Fock space
resolution of the irreducible module $L_\La$.  For $\widehat{sl(2)}$
(and $w=1$) this has been proved in \cite{BeFe}.

\section{BRST structure of $G/H$ coset theories}
\setcounter{equation}{0}

{}From the point of view of representation theory for affine Kac-Moody
algebras, the problem of constructing a coset model can be formulated
as follows:

Given a pair of algebras $(\wbg,\wbh)$, $\wbh\subset\wbg$, and a
corresponding pair of weights  $(\La,\la)$ (not necessarily
integrable), construct the coset module $L^{G/H}_{\La,\la}$ determined
by the decomposition of $L^G_{\La}$ into irreducible
$\wbh$-modules\footnote{Clearly one may often leave off the
superscripts $G$, $H$, and $G/H$ without causing confusion.  We will do
that whenever possible, and conversely, where it is essential we will
include them.} \eqn{decomp}{L^G_{\La}\,\cong\,
\bigoplus_{\la}L^{G/H}_{\La,\la}\otimes L^H_\la\,.}

Although it is possible to proceed in such generality (see \eg
\cite{PBW}), we will assume here that the embedding of $\wbh$ in $\wbg$
arises from a regular embedding of $\bh$ in $\bg$, so that one can
identify the Cartan subalgebras of $\bh$ and $\bg$ and choose the set
of positive roots of $\bh$ to be a subset of those in $\bg$.  Also, for
regular embeddings, the Dynkin embedding index  of $\bh$ in $\bg$ is
$j=1$.  It is then clear that any highest weight $\wbg$-module in the
category $\cal O$ is at least locally $\wbn_+^H$ finite as an
$\wbh$-module (in cases which are relevant for applications these
modules are in fact in the category $\cal O$ of $\wbh$-modules).

Recall that for the standard $G/H$ coset models \cite{GKO}, which will
be our main interest here, $\La \in P_+^{G,k}$, while the sum in
(\ref{decomp}) runs over  $\la \in P_+^{H,k}$.  In that case we may in
fact identify $L_{\La,\la}$ with the set of $\wbh$-singular states in
$L^G_\La$ at weight $\la$, and write down in terms of (semi-infinite)
cohomology \eqn{csbc}{L_{\La,\la}\,\cong  \,H^{\infty/2+0}_{\la}
(\wbn_+^H, L^G_\La)\,.} (Note that for $\wbn_+^H$ the ordinary and the
semi-infinite cohomologies coincide.)

However mathematically interesting, the complex
$\cC^{\infty/2+\bullet}_{\la}(\wbn_+^H, L^G_\La)$ is not very
attractive for a physicist  as it only has states with positive ghost
numbers, and thus  it cannot arise in a covariant manner from any
conformal field theory. Instead, the complexes we want to concentrate
on can be derived in the framework of the BRST quantization of gauged
WZW-models \cite{KaScht}, and correspond to the constraints $\wbh\sim
0$ on a larger module $L^G_\La\otimes V_{\la'}$, where $V_{\la'}$ is a
suitable highest weight $\wbh$-module with highest weight $\la'$
determined in terms of $\la$.  Then the space of physical states is
computed as the cohomology classes
\eqn{csbrst}{H^{\infty/2+n}(\wbh,L^G_\La\otimes V_{\la'})\,,\quad
n\in\ZZ\,.} The level  $k'$ of the affine weight $\la' $ must be $k' =
-k-2h^\vee$  (where $h^\vee$ is the dual Coxeter number of $\bh$), so
that the total central charge of $\wbh$ -- including the ghosts
contribution --  vanishes, and thus the BRST operator is nilpotent
\cite{KaScht}. The problem is to determine the appropriate module
$V_{\la'}$, and to compute the resulting cohomology.  For ease of
presentation we restrict our attention to the relative cohomology.  The
absolute cohomology follows in the usual way \cite{FGZ}.

\subsection{Cohomology of the BRST complex of the coset models}

For a given coset model, the BRST operator corresponding to the
differential of the complex which computes (\ref{csbrst}) may be
written explicitly in terms of the currents as \cite{KaScht}
\eqn{cosbrst}{d=:\oint\sum_{a=1}^{{\rm dim\,} {\bf h}}
c^a(z)[J_a(z)+\half J^{gh}_a(z)]:\,,} where  the normal ordering is
with respect to the $SL(2,\RR)$ vacuum, and the $J_a$ are
$\wbh$-currents on the product module. Since the ghost and anti-ghost
fields have conformal weights $0$ and $1$, respectively, the physical
vacuum $\om_0$ differs from the $SL(2,\RR)$ vacuum $|0\rangle_{gh}$ in
the ghost sector by the zero modes of the $c^a(z)$  in $\bn_+^H$, and
thus the weight of $\om_0$ is equal to
$\sum_{\al\in\De_+^H}\al=2\rh^H$.

{}From the decomposition (\ref{decomp}) it is clear that
\eqn{cohsplit}{ \scm\bull(\wbh,\wbt;L^G_\La\otimes
V_{\la'})\,\cong\,\bigoplus_\la L_{\La,\la} \otimes
\scm\bull(\wbh,\wbt;L^H_\la\otimes V_{\la'})\,.} Thus one only needs to
choose an appropriate $\wbh$-module $V_{\la'}$, and compute the
cohomology of $L_\la\otimes V_{\la'}$ for irreducible $\wbh$-modules
$L_\la$.  In particular one would like to understand whether for some
$V_{\la'}$ the sum on the r.h.s.\ collapses to just one term, as in
such a case the l.h.s.\ would give us a cohomological description of
the coset module which, unlike (\ref{csbc}), is ``covariant'' with
respect to the subalgbera $\wbh$.  Recalling the reduction theorem in
Section 3.3, and the cohomologies (\ref{vercoh}) and (\ref{wakscohom}),
it is natural to seek such $V_{\la'}$'s among Wakimoto modules and/or
Verma modules. Obviously this choice is also natural from the point of
view of the gauged WZW-model.

We will now proceed to compute (\ref{cohsplit}) for
$V_{\la'}=F_{\la'}^w$ and $M_{\la'}$. First a preparatory result, which
is an obvious consequence of Theorem \ref{redthm}  and the cohomologies
(\ref{vercoh}) and (\ref{wakscohom}) above.  \setcounter{cohprw}{0}
\thm{cohprw}{Lemma} {For a tensor product, $F^w_\la\otimes
F^{ww_0}_{\la'}$ ($w_0$ is the longest element in the Weyl group,
$W^H$), of two oppositely twisted Wakimoto $\wbh$-modules with
arbitrary weights $\la$ and $\la'$, \eqn{ctwmfr}{\scm
n(\wbh,\wbt;F^w_\la\otimes F^{ww_0}_{\la'})\, \cong\,\de^{n,0}
\de_{-\la-2\rh,\la'}\ \CC\,.} Similarly, for the Verma modules we have
\eqn{ctwmffr}{\scm n(\wbh,\wbt;M_\la^\star\otimes
M_{\la'})\,\cong\,\de^{n,0} \de_{-\la-2\rh,\la'}\ \CC\,.}} [For
$\wbh=sl(N)$ and $w=1$,  the above result was already proved in
\cite{Frprc,Sonna,HuYu,Sonnb,Sonnc} using explicit realizations of both
Wakimoto modules.  One should note that in the present derivation one
only uses cohomologies of separate Wakimoto modules with respect to
complementary subgroups of $\wbh$.] By choosing a suitable resolution
of the irreducible module $L^H_\la$ we can now construct a double
complex, as discussed in Section 3.5, and using the above lemma
essentially read off by inspection the following results.
\setcounter{bottgen}{1} \thm{bottgen}{Theorem} {Let $\la$ be an
integrable  weight of the Kac-Moody algebra $\wbh$ at level $k$ and
$\wbn_+^w$ a twisted subalgebra of $\wbh$ corresponding to a fixed
$w\in W^H$. Then \begin{itemize} \item As a $\widehat \bt$-module
\eqn{xxxxxx}{H^{\infty/2+n}(\wbn_+^w,L_\la)\,\cong\,
 \bigoplus_{\{w'\in \widehat W\,|\, \ell_w(w')=n\}} \CC_{w'*\la}\,.}
\item The following projection holds \eqn{projec}{\scm
n(\wbh,\wbt;L_\la\otimes F^{ww_0}_{\la'}) \,\cong\, \bigoplus_{\{w'\in
\widehat W\,|\, \ell_w(w')=n\}} \de_{w'*\la,-\la'-2\rh}\ \CC\,.} \item
The same result  holds for the Verma module  $M_{\la'}$ (resp.\
contragradient Verma module $M_{\la'}^\star$) if one substitutes
$\wbn_+^w\rightarrow\wbn_-$ (resp.\ $\wbn_+$) and
$\ell_w(w')\rightarrow -\ell(w')$ (resp.\ $\ell(w'))$ in (\ref{xxxxxx})
and (\ref{projec}).  \end{itemize}}

We should note that the first part follows directly from the mere
existence of a resolution of $L^H_\la$ in terms of Wakimoto modules
(see Theorem \ref{brrrr}), and their cohomology.  It was first derived
in \cite{FFcmp} as a generalization, to twisted subalgebras, of the
Bott theorem (part 3 of the above theorem) for integrable
representations of Kac-Moody algebras \cite{GL,PS}. One may also
observe that for each $n$ the direct sum in (\ref{xxxxxx}) consists of
one term for $\bh=sl(2)$, and is infinite for higher rank algebras. On
the other hand the $\de$-function in (\ref{projec}) will always project
out zero or one term, the latter when $-\la'-2\rh$ is in the image of
an integrable weight $\la$ under the twisted action of some $w'$.

One can prove similar theorems in the case of admissible
representations where one should use resolutions as in (\ref{PBCr}).
We will leave this as an exercise for the reader. (The
$\widehat{sl}(2)$ case has been discussed in \cite{Sonnb}.)

Finally by putting everything together we may summarize the general
case of the BRST cohomology for $G/H$ models in the following theorem:
\setcounter{cosgensum}{2} \thm{cosgensum}{Theorem}{Consider a pair of
Kac-Moody algebras $(\wbg , \wbh)$, an integrable weight $\La\in
P_+^{G,k}$, and an arbitrary weight $\la'\in P^{H,k'}$, where
$k'=-k-2h^\vee$. Then \begin{itemize} \item $H^{\infty/2+n}(\wbh,\wbt;
L_\La^G\otimes F_{\la'}^{ww_0})$ is always zero or one-dimensional, and
can be nontrivial for at most one $n$, \eqn{projecc}{\scm
n(\wbh,\wbt;L_\La^G\otimes F^{ww_0}_{\la'}) = \bigoplus_{\{w'\in
\widehat W\,|\, \ell_w(w')=n\}} L_{\La,-(w')^{-1}*(\la')-2\rh}\,.}
\item In particular, if we take $\la'=-\la-2\rh$, where $\la$ is an
integrable weight, $\la\in P_+^{H,k}$, we have \eqn{realproj}{\scm
n(\wbh,\wbt;L_\La^G\otimes F^{ww_0}_{-\la-2\rho }) = \de^{n,0}
L_{\La,\la}\,.} \end{itemize}}

\subsection{Free field resolutions of the coset models}

The last theorem above provides the crucial step towards the
construction of a resolution of the coset module, $L^{G/H}_{\La,\la'}$,
in terms of more manageable modules --  one must now simply replace
 the irreducible $\wbg$ module $L_\La^G$ in a controlled way.  Here we will
show how this is done, and examine the consequences.
 Let $({\cal R}_\La,\de)$ be a resolution of $L_\La$ in terms of
modules in the category $\cal O$, \eg in terms of Verma modules as in
Theorem \ref{bla} or in terms of Wakimoto modules as in Theorem
\ref{brrrr}. Consider the double complex $(\cC_{\La,\la},d_{rel},d')$
with spaces \eqn{dblcst}{\cC^{m,n}_{\La,\la'} \cong {\cal
R}^{m}_\La\otimes F^{ww_0}_{\la'}\otimes \cF_{gh,rel}^n\,,} and with
differentials $d=d_{rel}$  (the BRST operator of the relative
cohomology) and $d'=(-1)^{gh}\de$, where the prefactor counting the
number of ghosts is introduced so that $d$ and $d'$ anticommute. Let
$D=d+d'$ be the total differential in this complex.

\setcounter{oldstuff}{3} \thm{oldstuff}{Theorem}{For a pair of
Kac-Moody algberas $(\wbg,\wbh)$ and a corresponding pair of integrable
weights $(\La,\la)$, the cohomology of the complex
$(\cC_{\La,-\la-2\rh},D)$ defined above is
\eqn{cosrrs}{H^n(D,\cC_{\La,-\la-2\rh})\cong \delta^{n,0}
L_{\La,\la}\,.}}

\proof The proof follows from the first spectral sequence  in Section
3.2 and Theorem \ref{cosgensum} above -- simply note that the
$E_1$-term given in (3.13) has cohomology given by (\ref{realproj}),
and thus the spectral sequence collapses at $E_2$.$\Box$

In general one must be careful here in which order the cohomologies of
$d$ and $d'$ are computed, as the other spectral sequence corresponding
to this double complex may not collapse at the second term!  It does,
however, collapse if one chooses the resolution of $L_\La$ to be in
terms of Wakimoto modules of $\wbg$ that, as $\wbh$-modules, are
oppositely twisted to $F^{ww_0}_{\la'}$; \ie the subalgebras
$\wbn_+^{w(-)}$ and $\wbn_+^{(+)}$ of $\wbh$ act freely and cofreely,
respectively.  Thus, in this case one has further reduced the problem,
so that the resolution is via a complex whose spaces are just the
cohomology spaces $H^{\infty/2+\bull}( d,{\cal R}^n\otimes
F_{\la'}^{ww_0})$, and whose differential is just that induced from
$\de$ on these spaces.

When the resolution of $L^G_\La$ is in terms of Wakimoto modules, we
will call the complex of Theorem \ref{oldstuff} a free field resolution
of the $G/H$ coset model.

Of course there are other ways of splitting the differential $D$ in
this complex. In particular, one may try to eliminate the second
Wakimoto module by using the reduction theorem.  This can easily be
achieved by introducing $f_w$-degree. The 0-th order differential with
respect to $f_w$-degree is $D_0=d'+d_++d_-$, where $d_\pm$ are the BRST
charges corresponding to $\wbn_+^w$ and $\wbn_-^w=\wbn_+^{ww_0}$,
respectively, precisely as in Section 3.3.  Using the reduction theorem
we obtain then a smaller complex, $(\cC_{(red)\La,\la},d_+,d')$, in
which \eqn{redcmp}{ \cC_{(red)\La,\la}^{m,n}\cong ({\cal
R}^m_{\La}\otimes \cF_{gh,+}^n)_\la\,,} where $(\cdot)_\la$ denotes the
projection onto the subspace with the weight $\la$. Obviously, the
cohomology of this reduced complex also yields the same coset module.

If we carried out the same reduction procedure starting from a
resolution of the coset module, but with $M_{\la'}$ rather than
$F^{ww_0}_{\la'}$, we would arrive at the original ``free field''
realization of the coset model as  constructed in \cite{BMPnp}.

In the latter case one may even further reduce the complex by
evaluating the BRST cohomology of $d_+$.
At ghost number equal to zero it clearly consists of the subspaces
${\cal S}_{\La,\la}^m$ of singular vectors in ${\cal R}^{m}$.
Although it is not clear whether
the spectral sequence (whose first term we have just computed)
converges at the second term,\footnote{Note that we have changed the
order in which the cohomologies of $d_+$ and $d'$ are computed.} one
can give a separate (and rather subtle) argument which proves that the
complex $({\cal S}_{\La,\la},\de)$ yields a resolution of the coset
module.  Note that this complex is just a naive extension of
(\ref{csbc}), and it is the change of order in which two cohomologies are
evaluated that introduces the complication which must be dealt with
separately.  We refer the reader to \cite{BMPnp} and \cite{BMPrev} for
a more complete discussion.

\section{BRST cohomology of 2d gravity}
\setcounter{equation}{0}

In this section we will show how the results of the previous sections,
in particular the reduction theorem, can be applied to compute physical
states for $2d$ gravity coupled to $c\leq1$ conformal matter.  The
relevant algebra in this case is $\mb g = \pbv$, while the interesting
modules are the Virasoro Verma modules $M_{\De,c}$, irreducible modules
$L_{\De,c}$  and the so-called Feigin-Fuchs (or Fock space) modules
$F_{p,Q}$. Verma modules and irreducible modules are both labelled by
their conformal dimension $\De$, \ie $L_0$ eigenvalue of the highest
weight state, while the Feigin-Fuchs modules are labelled by the
momentum $p$ and background charge $Q$ of the free scalar field (we
refer to \cite{BMPcmpt} for any required clarification of the
conventions).  While the gravity sector will always be represented by a
Feigin-Fuchs module $F_{p^L,Q^L}$, for the matter sector one has the
choice of taking a Feigin-Fuchs module $F_{p^M,Q^M}$ (gravity coupled
to a free scalar field, \ie the two-dimensional string), or an
irreducible module $L_{\De^M,c^M}$ (gravity coupled to a minimal
model).

The crucial difference with the affine Kac-Moody case is that, contrary
to the Wakimoto modules, Feigin-Fuchs modules in general do not have a
simple cohomology due to the fact that they are neither free nor
co-free over part of the Virasoro algebra. As a consequence one finds
that the cohomology $H(\pbv,\pbv_0; F_{p^M,Q^M}\otimes F_{p^L,Q^L})$ is
already nontrivial, \ie discrete states occur, contrary to what we have
seen for the analogous cohomology in the affine Kac-Moody case.
Nevertheless, some of the techniques of the previous sections can be
applied by using specific properties of Feigin-Fuchs modules. For
instance, by making use of the fact that $c\geq25$ Fock modules
$F_{p,Q}$ are isomorphic to either the Verma module $M_{\De,c}$ or the
contragradient Verma module $M_{\De,c}^*$, depending on whether
$\eta(p) = \sign(i(p-Q))$ is positive or negative, respectively
\cite{Frplb}, we find by applying the reduction theorem \eqn{PBaa}{
H^{n}(\pbv,\pbv_0;  L_{\De^M,c^M}\otimes F_{p^L,Q^L}) \cong
H^{n}(\pbv_{\eta(p^L)},L_{\De^M,c^M})_{1-\De(p^L)} } In order to
compute $H^{n}(\pbv_{\eta(p^L)},L_{\De^M,c^M})$, we take a resolution
of $L_{\De^M,c^M}$ in terms of (contragradient) Verma modules
\cite{FeiginFuchs}, depending on $\eta(p^L)$, and proceed as outlined
in Section 3.  The result can be summarized as follows
\cite{LZclo,LZceo,BMPcmpt,LZcmpt,BMPcar}

\thm{cohtwod}{Theorem}{ Let $({\cal R}^{(n)},d')$ be a resolution of
$L_{\De,c}$, where each ${\cal R}^{(n)}$ is a direct sum of Verma
modules, contragradient Verma modules or Feigin-Fuchs modules. Let
$\De^{(n)}$ be the conformal dimension of any one of the modules in
${\cal R}^{(n)}$.  Denote $\tilde{\Ep}(\De,c) = \{ 1-\De^{(n)} \}$. For
$\De=1-\De^{(n)}$, for some $n$, define $d(\De)=|n|$ (these definitions
for $\tilde{\Ep}(\De,c)$ and $d(\De)$ do not depend on the specific
resolution).  Then we have \begin{enumerate} \item $
H(\pbv,\pbv_0;L_{\De^M,c^M}\otimes F_{p^L,Q^L}) \neq 0 \qquad {\rm
iff}\qquad \De(p^L) \in \tilde{\Ep}(\De^M,c^M)\,.  $ \item For
$\De(p^L) \in \tilde{\Ep}(\De^M,c^M)$ we have \eqn{vermacohomgra}{ {\rm
dim}\ H^{n}(\pbv,\pbv_0;L_{\De^M,c^M}\otimes F_{p^L,Q^L}) =
\de_{n,\eta(p^L)d(\De(p^L))}\,.} \end{enumerate}}

\section{Concluding remarks}

We have shown in these lectures how various methods of homological
algebra can be applied to compute BRST cohomologies that arise in
conformal field theory. The central role here is played by a reduction
theorem which, roughly speaking, allows one to project out subspaces of
a module by first tensoring it with a suitable highest weight module
and the computing the BRST cohomology of the product module.
 In particular, we have used this technique to construct free field
resolutions of $G/H$ coset theories, by imposing the BRST operator on
the usual free field resolution of the WZW-model of $G$ tensored with
an appropriate Fock space of the free field realization of the
WZW-model of $H$.  These new resolutions of coset theories should allow
one to repeat the same steps that have been carried out  in the case of
the usual (\ie ungauged) WZW-models (see \eg \cite{BMPrev} and the
references therein).  The most important outstanding problem in this
direction is the construction of screened vertex operators, and the
computation of the fusion rules.

\end{document}